\documentclass[twocolumn,english,aps,prb,noshowpacs,superscriptaddress,amssymb,amsfonts]{revtex4}
\usepackage[T1]{fontenc}
\usepackage[latin9]{inputenc}
\usepackage{amsmath}
\usepackage{epstopdf} 
\usepackage{graphicx}
\usepackage{amssymb}
\usepackage{color}

\makeatletter
\newcommand{\be}{\begin{equation}}
\newcommand{\ee}{\end{equation}}
\newcommand{\bea}{\begin{eqnarray}}
\newcommand{\eea}{\end{eqnarray}}

\newcommand{\sz}{\sigma^{z}}
\newcommand{\sx}{\sigma^{x}}
\newcommand{\mz}{\mu^{z}}
\newcommand{\mx}{\mu^{x}}
\makeatother

\usepackage{babel}

\makeatother

\usepackage{babel}

\begin{document}

\title{Emergent  Space-time Supersymmetry at the Boundary of a  Topological Phase}
\author{Tarun Grover}

\affiliation{Kavli Institute for Theoretical Physics, University of California, Santa Barbara, CA 93106, USA}
\author{D. N. Sheng}
\affiliation{Department of Physics and Astronomy, California State University,
North Ridge, CA 91330, USA}
\author{Ashvin Vishwanath}
\affiliation{Department of Physics, University of California,
Berkeley, CA 94720,
USA}

\begin{abstract}

In contrast to ordinary symmetries, supersymmetry interchanges bosons and fermions. Originally proposed as a symmetry of 
our universe, it still awaits experimental verification. Here we theoretically show that supersymmetry emerges naturally in topological 
superconductors, which are well-known condensed matter systems. Specifically, we argue that the quantum phase transitions at the 
boundary of topological superconductors in both two and three dimensions display supersymmetry when probed at long distances and times.    Supersymmetry entails several experimental consequences for these systems, such as, exact relations between 
 quantities measured in disparate experiments, and in some cases, exact knowledge of the universal critical exponents. The topological 
 surface states themselves may be interpreted as arising from spontaneously broken supersymmetry, indicating a deep relation between 
 topological phases and SUSY. We discuss prospects for experimental realization in films of superfluid  He$_3$-B.

\end{abstract}

\maketitle
\section{Introduction}
%\tableofcontents

In the 1970s, space-time ``supersymmetry'' (SUSY), was proposed as a possible invariance of our universe \cite{susy_origin1,susy_origin3}. Unlike any other symmetry, SUSY interchanges bosons and fermions and when  applied twice, generate translations of space and time, which ultimately leads to the conservation of momentum and energy \cite{noether}. SUSY theories were actively pursued to attack long-standing problems such as the hierarchy problem in the elementary particle physics \cite{susy_appl}, but despite sustained effort, it has yet to be experimentally established in Nature.

\begin{figure}
\begin{centering}

\includegraphics[scale=0.30]{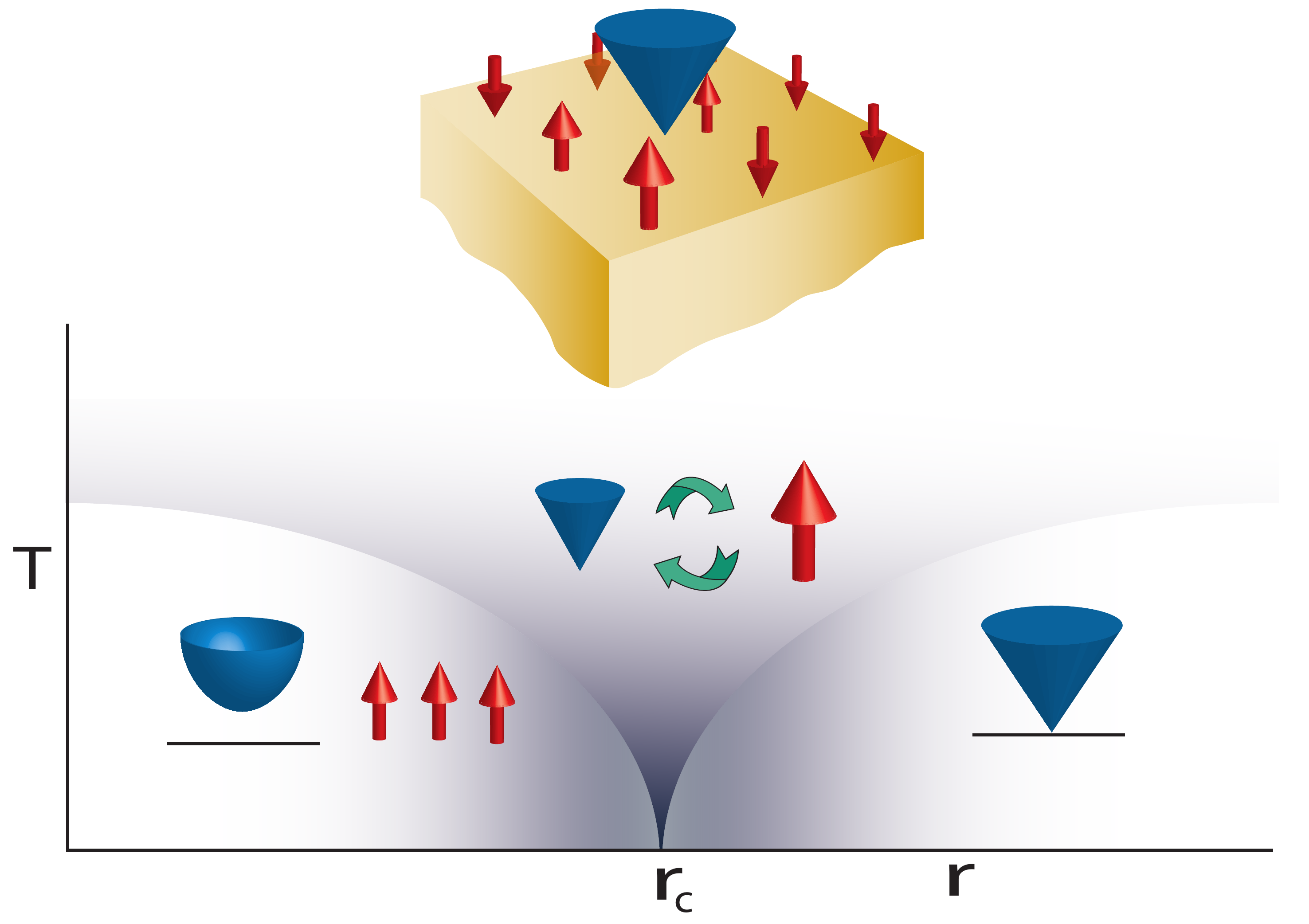}
\par\end{centering}
\caption{Phase diagram of a three dimensional topological superconductor, as Ising magnetic fluctuations (denoted by red arrows) at the boundary couple to the Majorana fermions (blue cone). When the tuning parameter $r < r_c$, the Ising spins order leading to a gap for the Majorana fermions. In the main text, it is argued that the critical point that separates the two sides is supersymmetric, where bosons (Ising order parameter) and Majorana fermions transform into each other. Similar phase diagram is obtained for two-dimensional topological superconductors (Fig.\ref{fig:phasedia2d}).} \label{fig:phasedia3d}
\end{figure}

  In this paper we theoretically show that certain condensed matter systems display the remarkable phenomena of \textit{emergent supersymmetry}, that is, space-time SUSY naturally emerges as an accurate description of these systems at low energy and at long distances, although the microscopic ingredients are not supersymmetric.     The physical systems we mainly consider are topological superconductors  \cite{roy2008, qi2008}, in which fermions, which may be electrons or fermionic atoms such as He-3, pair together in a special way. The resulting state has an 
  energy gap to fermions in the bulk but gapless excitations at the surface.  The surface excitations are prohibited from acquiring a gap 
  due to time reversal symmetry. We consider the spontaneous symmetry breaking quantum phase transition at the surface, using 
  numerical and analytical techniques, and establish emergent supersymmetry in both two and three dimensions (Fig.\ref{fig:phasedia3d}). 
  
Our result is similar in spirit to Friedan, Qiu, Shenker \cite{friedan1}, who showed that the 1+1 dimensional tricritical Ising model, which can be
accessed by tuning two 
parameters, is supersymmetric. Few other proposals that realize SUSY by tuning two or more 
parameters have been made as well  \cite{kunyang,liza2}.  Here, we require that  SUSY  be achievable by tuning 
only a single parameter, and separates two distinct phases, akin to a conventional quantum critical point \cite{Sachdev,footnote1}. This  is crucial for our results to be experimentally 
realizable. We also require that our theory have full space-time SUSY rather than only a limited `quantum-mechanical' SUSY \cite{footnote_qmsusy}. Furthermore, in contrast to the strategy adopted in \cite{friedan1}, our approach is not restricted to 1+1 dimensional theories. This automatically 
ensures translation invariance in space and time, and will lead to experimental consequences, as we discuss below.

There has been an explosion of activity in the field of topological phases since the discovery of $\mathbb{Z}_2$ topological insulators (TIs)\cite{TIreview1,TIreview2,zhang2007,hseih2008}. We will focus on a set of closely related phases, the time-reversal invariant topological superconductors \cite{roy2008,qi2008} (TSc) --- which include the well known B-phase of superfluid He-3 \cite{He3}. These phases exist in both two and three spatial dimensions \cite{schnyder2008,kitaev2009} and support Majorana modes at their boundary, which are protected  by time-reversal symmetry from acquiring an energy gap. Spontaneous breaking of this symmetry
provides a natural mechanism to gap them out. For example, electron-electron interactions at the surface could lead to magnetic order, which breaks time reversal symmetry. A natural question is: how do  the edge modes evolve as the magnetic order sets in? Surprisingly, we will see that space-time SUSY naturally emerges at the onset of magnetic order.

\begin{figure}
\begin{centering}
%\vspace{-0.8in}
%\hspace{-0.8in}
\includegraphics[scale=0.41]{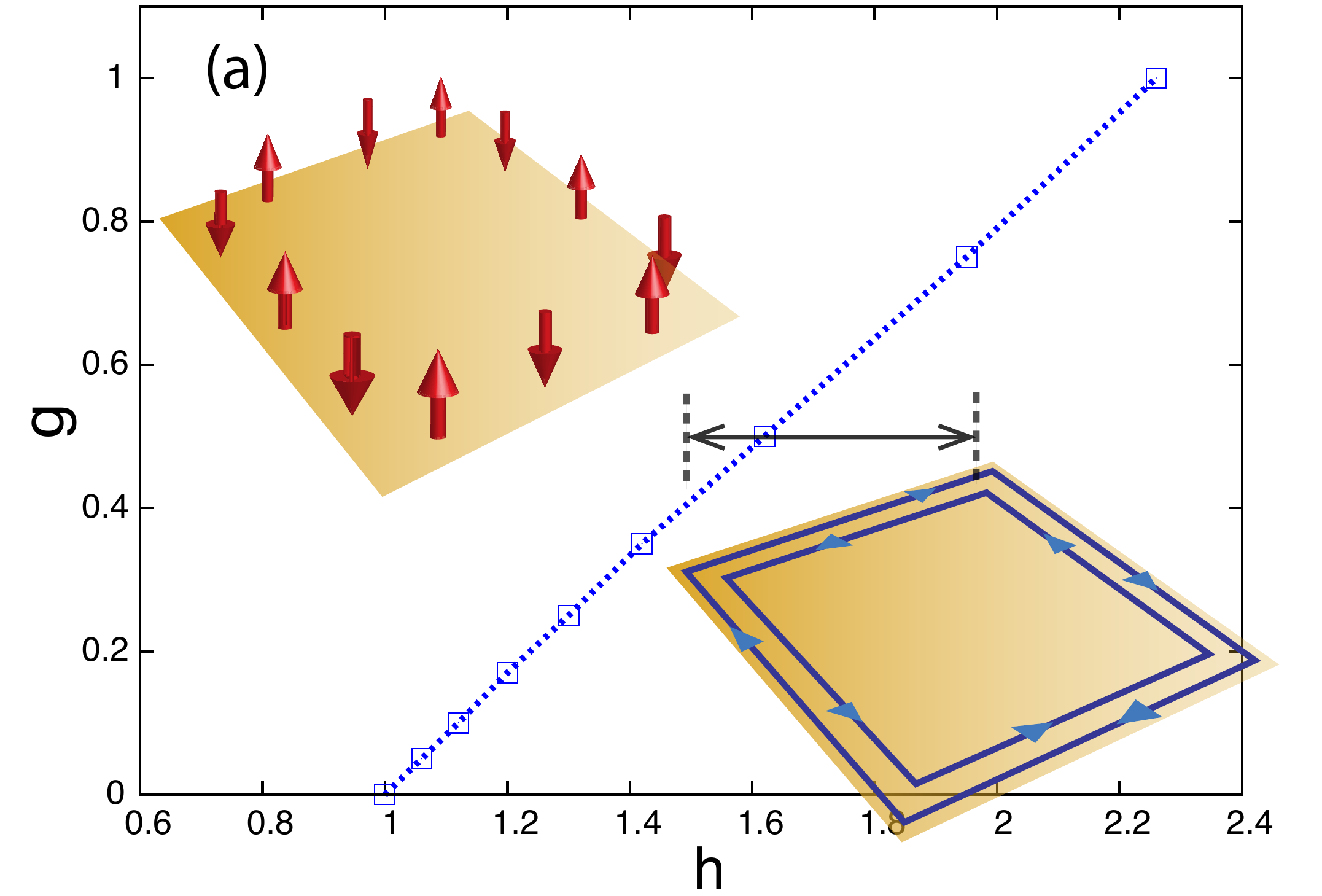}
\includegraphics[scale=0.45]{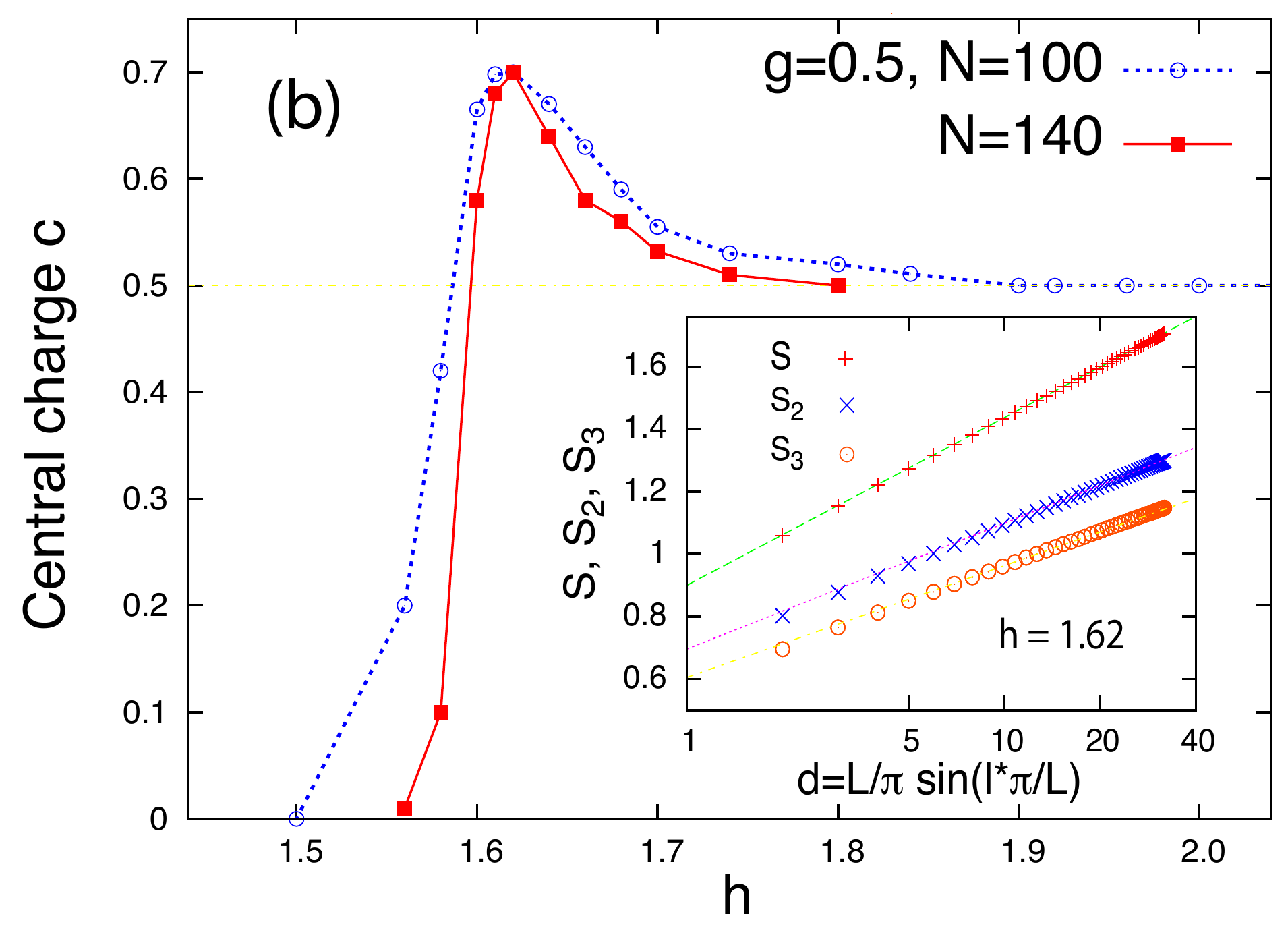}
\par\end{centering}
\caption{(a) The phase diagram of the Hamiltonian $H$ in Eqn.\ref{eq:lattice1d}, which realizes the Majorana edge of a two dimensional time-reversal invariant topological superconductor coupled to Ising magnetic fluctuations.  The vertical axis $g$ is the coupling between the Majorana fermions and the Ising order parameter. At large $h$, the Ising spins  disorder and the counter-propagating Majorana modes remain gapless. As $h$ decreases, the ordering of Ising spins leads to a gap for the Majorana modes. The black arrows indicate the region of the phase diagram detailed in Fig.\ref{fig:phasedia2d}(b).  (b) Central charge $c$ as a function of $h$, for fixed $g = 0.5$. For $h > h_c\,(=1.62)$, one finds $c = 1/2$, consistent with gapless Majorana modes whole for $h < h_c $, one finds $c = 0$ indicating the gapped phase. The critical point separating the two phases is characterized by $c = 7/10$ which corresponds to supersymmetric tri-critical Ising point.  The inset shows the von Neumann entropy $S$ and the Renyi entropies $S_2, S_3$ at the critical point, which were used to deduce the central charge.
} \label{fig:phasedia2d}
\end{figure}

\section{D=1+1 Emergent SUSY at the Boundary of a TSc}
The D=2+1 dimensional topological superconductor, protected by the time reversal symmetry, provides the simplest setting to address this question. While the bulk of the superconductor is gapped, the boundary, a D$_{\rm edge}$ =1+1 dimensional system,  contains a pair of Majorana modes $\chi_R, \chi_L$ that propagate in opposite directions. The aforementioned instability of the edge may be described by introducing an Ising field $\phi$ that changes sign under time-reversal. The action for the full theory is given by
\begin{eqnarray}
S^{d+1} &=& \int d\tau \,d^dx\, \,  [ \frac12 \bar{\chi} \displaystyle{\not}\partial \chi  +\frac12 (\partial_\tau \phi)^2 + \frac{v_\phi^2}{2} (\nabla \phi)^2 \nonumber \\ & & +\frac r2 \phi^2 + g\phi \bar{\chi}\chi  +u\phi^4] \label{eq:fieldtheory}
\end{eqnarray}

\noindent with $d=1$, $\chi = [\chi_R\,\, \chi_L]^T$, and we have employed the conventional Dirac gamma matrices for the relativistic fermion $\chi$ (e.g., in terms of Pauli matrices  $\gamma_0=\sigma_y,\,\gamma_1=\sigma_x$). The broken symmetry phase is characterized by  $\langle \phi \rangle \neq 0$, which leads to a mass gap $g\langle \phi \rangle $ for the fermions.

The mode count in the action  $S$ is favorable for  $\mathcal{N}=1$ SUSY in D=1+1 \cite{wessbagger}, with the bosons 
$\phi$ and Majorana fermions $\chi$ as super partners. We now show that this is indeed the case at the critical point using a numerical simulation of a D = 1+ 1 
lattice model that reproduces the action in Eqn.\ref{eq:fieldtheory}  at low energies. The model is given by 

\begin{eqnarray}
\label{H_simulate}
\label{eq:lattice1d}
H &=&  -i\sum_j \left [ 1-g\mu^z_{j+\frac12}\right ]  \chi_j \chi_{j+1} + H_b \\ \nonumber
H_b &=& \sum_j   \left [{\mathcal J}\mu^z_{j-1/2}\mu^z_{j+1/2} - h  \mu^x_{j+1/2}\right ] 
%H_{\rm int} &=&  ig \sum_j \chi_j \mu^z_{j+\frac12}\chi_{j+1}\\ \nonumber
\end{eqnarray}
Here $\chi_j$ is a single Majorana fermion at site $j$ while the Ising spins $\mu^z_{j+1/2}$ sit on bond centers. When $h \gg1$, $\langle \mu^z \rangle = 0$ and  lattice translation symmetry ensures that the Majorana fermions are gapless. As $h$ decreases, at some point, $\mu^z$ orders antiferromagnetically leading to a mass gap for the fermions through the coupling $g$, reproducing the field theory in Eqn.\ref{eq:fieldtheory} at and near the critical point. $\mathcal J$ tunes the relative bare velocities between the boson and the fermion modes, similar to $v_\phi$ in Eqn.\ref{eq:fieldtheory}.

We numerically simulate a spin version of Eqn.\ref{H_simulate} using the  Density Matrix Renormalization Group (DMRG)   
method \cite{white1992}. See Appendix \ref{sec:1dmodeldetails} for details of the model. We first set ${\mathcal J} = 1$, and determine the phase diagram as a function of the couplings $g,\,h$. A key tool to characterize phases is the central charge $c$, which is a measure of the number of gapless modes and directly accessed by DMRG. While the gapped edge has $c=0$ and the Majorana edge has $c=1/2$, the simplest supersymmetric theory, the tricritcal Ising model is expected to have $c=7/10$.    % 

\subsubsection{Results of Numerical Simulation}
Fig.\ref{fig:phasedia2d} shows the numerically determined phase diagram. At larger $h$, a gapless phase is obtained  that is separated by a critical line from a fully gapped ordered phase at small $h$. To characterize the critical theory, consider crossing the phase boundary along fixed $g$, say, $g = 0.5$, while monitoring the central charge $c$ (Fig.\ref{fig:phasedia2d}b).  At  small $h$, $c\approx 0$ indicating a gapped phase.  At $h>h_c \approx 1.62$, the central charge saturates at  $c \approx 0.5$, indicating that the symmetry is restored and a gapless Majorana mode is present.  At the transition, $h=h_c$, we find  $c \approx 0.7$.   The jump from $c = 0$ to $ c= 0.7$ becomes sharper as the system size increases from $L=100$ to $140$ with the decrease of  the finite size effect. The excellent fit shown in the inset of Fig.\ref{fig:phasedia2d}b, where the entanglement entropy for different sized partitions of the system tuned to the critical point is compared with the form expected for $c=7/10$, confirms that \textit{the phase transition lies in the tricritical Ising universality, which is known to be supersymmetric} \cite{friedan1}. The correlation functions of various local operators also confirm that this statement (see Appendix \ref{sec:tricritcorr}). Note that we access this transition by tuning a {\em single} variable.

\begin{figure}
\begin{centering}
\vspace{-0.5in}
\hspace{-1in}
\includegraphics[scale=0.43]{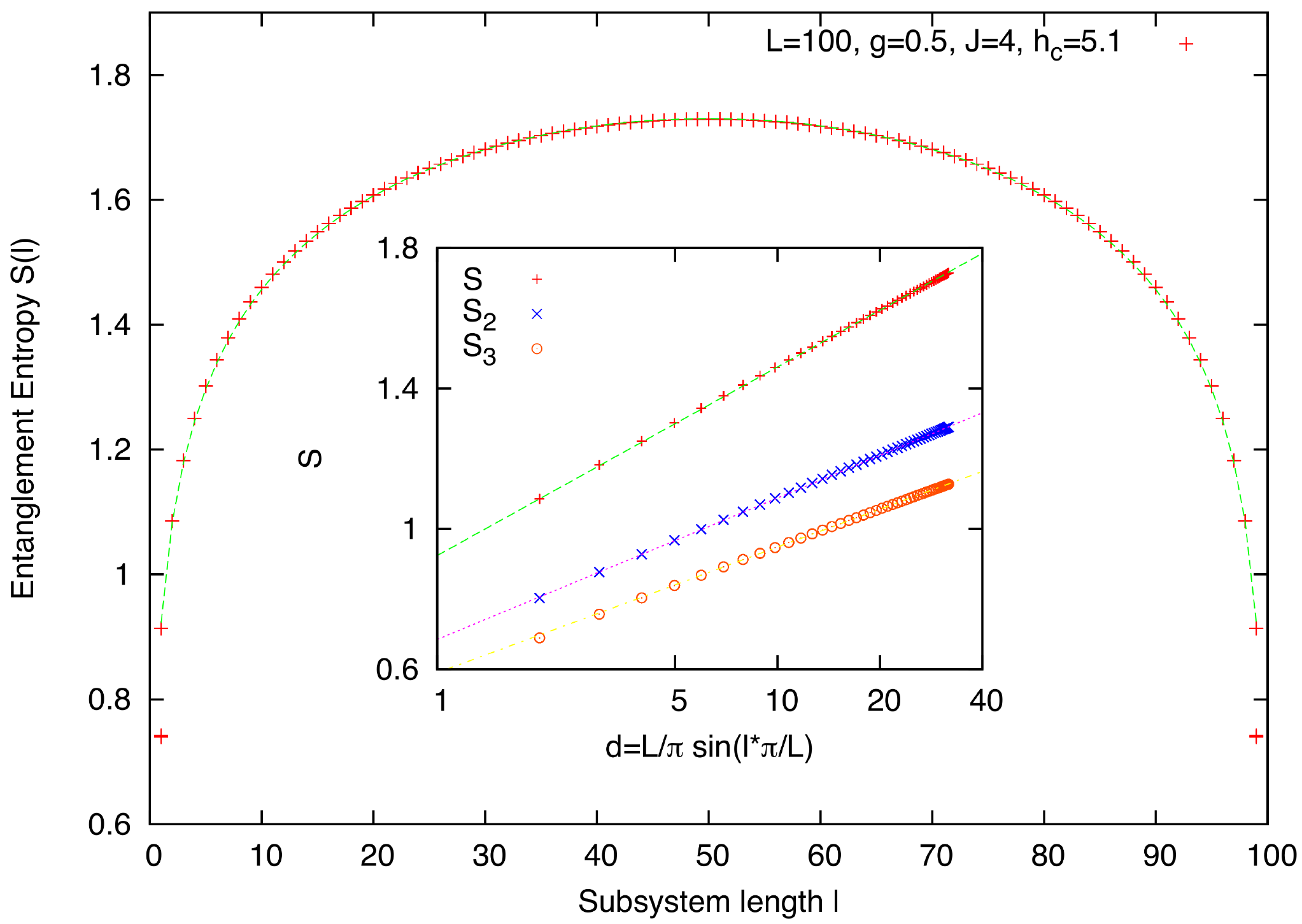}
\par\end{centering}
\caption{Entanglement entropy at the critical point for the 1+1-D lattice model for the parameter $\mathcal{J} = 4$. The paramater $\mathcal{J}$ equals the ratio of the bare verlocity of the Majorana fermion to that of the boson $\phi$. The above curve shows that the supersymmetric critical point with central charge $c= 7/10$ survives even when the velocity anisotropy is four. The red crosses are the numerical data while the green curve is the theoretical expected result for central charge $c = 7/10$. The inset shows the Renyi entropies $S_n$ which also fit perfectly to $c = 7/10$.} \label{sup1}
\end{figure}

{\em Velocity Anisotropy:} We consider introducing an asymmetry in the bare velocities of boson and Majorana modes. This is done by varying  the parameter $\mathcal{J}$ between $\mathcal J = 1$ (nearly equal bare velocities), and $\mathcal J = 4$ (different bare velocities) , we obtain a second order transition in the same universality class (see Figure \ref{sup1}). This implies that even when the velocities of boson and fermion modes differ at the lattice scale, they ultimately converge to a single velocity at low energies.

\section{D=2+1 SUSY at the surface of a TSc}

Now consider a 3+1-D topological superconductors, whose two dimensional surface supports gapless Majorana fermions, protected by time reversal symmetry. As advertised earlier, we study spontaneous breaking of time reversal symmetry that gaps out the surface states. In the absence of numerics, we resort to an analytical calculation within an $\epsilon$ expansion to unravel the nature of quantum criticality and argue that 2+1 dimensional SUSY arises here.

The relevant action is again given by Eqn.\ref{eq:fieldtheory} with $d=2$, where $\chi^T=(\chi_1,\,\chi_2)$ is a two component real fermion field with $\bar{\chi} = \chi^T\gamma^0 \,{\rm and}\, \displaystyle{\not}\partial =  \gamma^\mu\partial_\mu $ where, for example we can take $\gamma^0 = \sigma_y,\, \gamma^1=-\sigma_z,\,\gamma^2 = \sigma_x$ in terms of the Pauli matrices. At $r=0$, this action has ${\mathcal N}=1$ SUSY  
\cite{wessbagger} in 2+1D if one sets 
\be
u = \frac{g^2}2
\label{SUSYCoupling}
\ee and $v_\phi = 1$.
The SUSY transformation is implemented by a pair of real fermionic generators $\varepsilon$:
\begin{eqnarray}
\delta \phi &=& \bar{\varepsilon} \chi \\
\delta \chi  &=& [-\displaystyle{\not}\partial \phi + g \phi^2] \varepsilon 
\end{eqnarray} 
We ask if SUSY emerges in the low energy limit as we tune to the quantum critical point, \textit{without} enforcing these conditions microscopically?
The coupling $g$ between critical bosonic modes and free fermions is readily seen to be strongly relevant at the decoupled fixed point $g=0$. 
While the spinor structure is held fixed, the dimensionality is varied in the integrals as described in detail in Appendix \ref{sec:3Drg}. On coarse graining the theory to scales $a' = a(1+dl)$, the coupling constants flow at one loop order according to: 

\begin{eqnarray}
\frac{dG}{dl} &=& \epsilon G -14G^2 \\
\frac{dU}{dl}&=& \epsilon U -4GU -36U^2+4G^2
\end{eqnarray} 
where the coupling constants  $G = g^2N_D$ and $U=uN_D$, and $N_D$ is a constant (the volume of the $D$ dimensional sphere). The fixed point is reached at the couplings:
\begin{equation}
\frac{G^*}2=U^*=\epsilon/7
\end{equation}
which implies that the fixed point action to this order automatically satisfies Eqn.\ref{SUSYCoupling},  and $v_\phi = 1$, indicating an emergent SUSY at this quantum critical point. See Appendix \ref{sec:3Drg} for the details. Furthermore, consistent with SUSY \cite{wessbagger}, the scaling dimensions $\Delta_\chi, \Delta_\phi$ of $\chi, \phi$ satisfy $\Delta_\chi = \Delta_\phi + 1/2$  (with $\Delta_\phi = \frac12 + \frac{\epsilon}{14}$).  Setting $\epsilon=1$ provides an estimate of the critical exponents for the D=2+1 dimensional surface. While there are no exact results to compare this against, it is amusing to note that substituting $\epsilon=2$ to access the $1+1$ dimensional critical point yields a value of the anomalous exponent $\eta_\phi=\epsilon/7=0.3$ relatively close to the exact $\eta_{1+1}=0.4$. 

{\em Velocity Anisotropy:} At the fixed point the boson and fermion velocities are equal as demanded by Lorentz invariance. Consider introducing a small difference between these two velocities, $\Delta v = v_\phi - v_\chi$ which is much smaller than the mean velocity $v=(v_\phi + v_\chi)/2$. One finds (see Appendic \ref{Sec:SUSYvelocity}) the flow of this velocity anisotropy within the $\epsilon$ expansion to be: $\frac{d\,\Delta v}{dl} 
 \approx   -\frac{5 \epsilon}{21 v^3} \Delta v$, i.e. the velocities approach each other in the low energy limit. Thus equal velocities are recovered even if they are not tuned to equality in the bare theory.

\section{Towards an Experimental Realization}  

\begin{figure}[htbp]
\begin{center}
\includegraphics[scale=0.30]{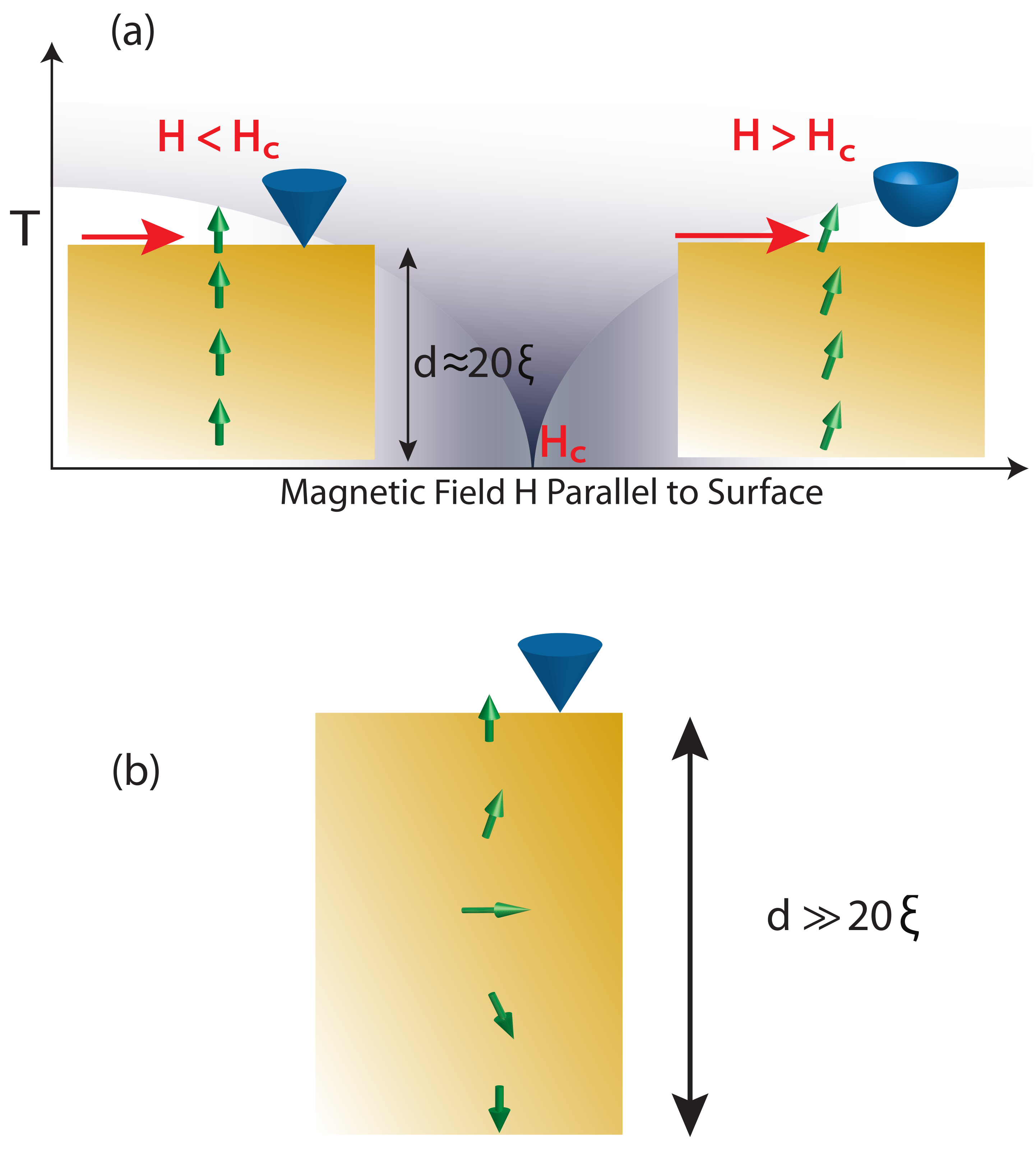}
%Fig. 1
\end{center}
\caption{(a) Mean-field phase diagram for a thin film of superfluid He$_3$-B as a magnetic field parallel to the surface is applied. The $\hat{\bf n}$ vector (green arrow) is oriented along the vertical direction, perpendicular to the surface. A weak field applied in plane does not open a gap to the majorana fermions on the boundary, since a residual symmetry, composed of 180 degree rotations and time reversal is present, that preserves the gapless surface state. However, on increasing the field above $H_c\approx 30 {\rm Gauss}$, the $\hat{\bf n}$ vector spontaneously tilts into the plane leading to a gap on the surface and hence a compensating energy gain. (b) For this transition to occur spontaneously on the two surfaces independently, one needs to consider films much thicker than the  $\hat{\bf n}$ healing length. Then, in the bulk the $\hat{\bf n}$ vector is pinned along the field. In principle this breaks symmetry, but since this is sufficiently far away from the surface is expected to have a negligible effect.
}
\label{He3}
\end{figure}

Consider a slab of He$_3$-B, with surfaces perpendicular to the $z$ direction (Fig.\ref{He3}).  We use the notation of Volovik \cite{volovikbook1} to write the single particle Hamiltonian:

\begin{equation}
H = \frac{p^2-p_F^2}{2m} \tau_z -\frac{\gamma}{2} {\bf H}\cdot{\bf \sigma} + \tau_x \sum_{\mu=x,y,z}\Delta_{\mu} \tilde{\sigma}^\mu \frac{p^\mu}{p_F}
\end{equation}

\noindent where the gap function $\Delta_{\mu} =\left (\Delta_\parallel,\Delta_\parallel,\Delta_\perp\right ) $ and the rotated Pauli matrices ${\bf \tilde {\sigma }} = {\mathcal R}{\bf \sigma}$, and ${\mathcal R}({\hat {\bf n},\varphi})$ is the order parameter of the superfluid, resulting from the breaking of relative rotations between orbital and spin directions. Weak dipolar interactions between the atoms in the cooper pairs pin this rotation matrix, parameterized in terms of the rotation axis $\hat {\bf n}$ and angle $\varphi$, to $\varphi = -\frac14 \frac{\Delta_\perp}{\Delta_\parallel}$. For a sufficiently thin sample with thickness $d<20\xi \sim 1.6\mu m$ (but not so thin that the B phase is itself destroyed in favor of the A phase, so $d>9-13\xi$, Ref.\cite{Sauls}), the $\hat {\bf n}$ vector is uniform in the sample. 
%If the full rotation symmetry were present, all directions of $\hat {\bf n}$ would be energetically equivalent. 
However, since the rotation symmetry is reduced within the slab geometry, a specific direction is selected. In fact the dipolar interactions pin this vector along the $z$ axis. 
Another mechanism to control the direction of $\hat {\bf n}$ appears from the Zeeman coupling to an external magnetic field. If the field is also along the $z$ axis, then, the $\hat {\bf n}$ vector continues to be pinned along this direction. The source of coupling arises from the surface states. An energy gap to surface states reduces their energy, the size of this gap being proportional to the projection $l_z = \hat{H}_\mu {\mathcal R_{\mu\nu}(\hat {\bf n},\varphi)}$. Now, if the field $H$ and $\hat {\bf n}$ are along $z$, this produces the maximum gap \cite{Mizushima,Zhang}. 

However, there is an interesting competition if the magnetic field is applied in plane i.e. $\hat{H} = \hat{x}$. Now, if $\hat {\bf n}$ is still along the $z$ axis, then this induces an $l_z=0$ and the surface states remain gapless. Physically this can be understood \cite{Mizushima} as resulting from a residual symmetry, rotation by 180 degrees and time reversal, which leaves the system invariant (note that time reversal followed by reflection in the x-z plane is not a symmetry since reflection changes the rotation angle $\varphi\rightarrow -\varphi$).  However, if the $\hat {\bf n}$ vector tips into the x-y plane, this symmetry is spontaneously broken and generates a mass for the Majorana fermions. If we write the $\hat {\bf n} = (n_x,\,n_y,\,n_z)$ then:

\begin{equation}
l_z = n_x n_z (1-\cos \varphi) +n_y \sin \varphi 
\end{equation}
when $n_z\approx 1$, this is maximized by components $(n_x,\,n_y)=\sigma (\sin \varphi/2,\,\cos \varphi/2)$ for which:
\begin{equation}
l_z \approx 2\sigma \sin \varphi/2
\end{equation}
Thus the surface Majorana fermions acquire a mass term $m=g \sigma$ where $g=2H\sin \varphi/2$. The energetic incentive for an in-plane component increases with the applied magnetic field. Thus, for fields $0\leq |H| < H_c$ we have a gapless 
symmetric state  while for $H_c<|H|$ spontaneous symmetry breaking occurs leading to a gap for surface Majorana modes.  In Ref.\cite{Mizushima}, the critical field was predicted to be $H>H_c=30$ Gauss, which is readily accessible.  The effective theory for the transition is:

\begin{eqnarray}
{\mathcal L} &=& L_\chi+L_\sigma + g\sigma \bar{\chi}\chi \\
L_f &=& \frac12 \bar{\chi} \gamma^\mu p_\mu \chi \\
L_\sigma &=& \frac12 (\partial_t \sigma)^2- \frac{c^2}{2}(\nabla \sigma)^2 -M^2 \sigma^2 - u \sigma^4
\end{eqnarray}

\noindent where we have set the velocity of the fermions to unity, and $c$ is the velocity of the bosonic modes, and $M$ is the gap set by the fact that $\sigma=0$ is preferred since the $\hat{\bf n}$ is oriented along $\hat{\bf z}$ direction in the absence of a magnetic field. This is identical to the action $S^{2+1}$ (Eqn.\ref{eq:fieldtheory}), which, as discussed above, flows to the $\mathcal{N} = 1$ SUSY fixed point within a $D = 4- \epsilon$ calculation.

We however note the challenges to realizing SUSY in this physical system. The low temperatures of the superfluid transition restricts the temperature window for experiments and the bare boson and fermion velocities are expected to be rather different, with a ratio controlled by $E_F/\Delta$. 
Furthermore, while the bare velocities of the bosonic mode is expected to be of order $v_{\rm Fermi}$, the velocity of the surface fermions is controlled by superfluid gap $\Delta$. For realizing SUSY, the difference between these two velocities must flow to zero at low energies, an alternative scenario being that the transition becomes first-order due to bare velocity difference.

Furthermore, to have a truly 2D transition, one  needs the order parameters on the two surfaces to fluctuate independently (see Fig.\ref{He3}b). Deep in the bulk, there is a weak pinning field on the $\hat{\bf n}$ vector, which leads to $H_{\rm pinning} = -\lambda [H\cdot \hat{\bf n}]^2$. Therefore the $\hat{\bf n}$  vector is aligned along the $\pm x$ direction in the bulk. It evolves from being along $z$ direction at the surface to being entirely along $\pm x$ in the bulk. In principle, this breaks the symmetry and produces a gap on the surface. However, when the film thickness $d \gg \xi$, the symmetry breaking effect at the surface is expected to be very small since the healing length of the $\hat{\bf n}$ order parameter is much longer than the coherence length.

\section{SUSY on the 3D Topological Insulator surface:} 

The surface states of a 3D topological insulator  \cite{TIreview1,TIreview2,TIreview3,zhang2007,hseih2008} consist of \textit{Dirac} fermions at a chemical potential $\mu: H = \sum_{k_x, k_y} c^{\dagger} \left(k_x \sigma_y - k_y \sigma_x - \mu \right) c  \label{eq:helicalTI3D}$. To realize SUSY, we consider fine-tuning $\mu$ to zero and consider the instability of these surface modes to an \textit{s-wave superconductor} \cite{grover2012,ponte}. This {\em multi critical } point can potentially be driven by intrinsic interactions, and may also be realized by patterning the surface with a Josephson junction array. We restrict ourselves to particle-hole symmetric superconducting fluctuations. The effective action for the coupled system near the  transition is given by \cite{grover2012,ponte}

\bea
S & = & \int d^3x \,\,\left[\overline{\psi} \displaystyle{\not}\partial \psi + g\,\left(\phi\,\, \psi^T \epsilon \psi  + c.c.\right) \right. \nonumber \\
& & + \left. |\partial_\tau \phi|^2 + c^2|\vec{\nabla} \phi|^2 + r|\phi|^2 + u |\phi|^4\right] \label{eq:susy3dTI}
\eea
where $\phi$ is the superconducting order parameter, and the fermion field $\psi^T = \left[c_{\uparrow} \,\,\,c_{\downarrow}\right]$,\, 
%$\{\gamma_0,\gamma_1,\gamma_2\} = \{\sigma_z, \sigma_x,\sigma_y \}$ 
and $\epsilon$ is the  two-dimensional antisymmetric tensor. The above action is known to flow to a supersymmetric fixed point \cite{balents1998, sungsik}. The SUSY corresponds
 to a $\mathcal{N} = 2$ Wess-Zumino model. Most remarkably, in this case SUSY allows one to calculate the exact anomalous dimensions of the boson and Fermi fields, with \cite{aharony} $
\eta_\phi= \eta_\psi= 1/3 $.

\section{Conclusions}
It is well known that the spontaneous breaking of SUSY leads to generation of a massless fermion, called the `goldstino' mode \cite{das1978}. An example is provided by the action in Eqn.\ref{eq:fieldtheory} with d=1, where the flow from the critical point ($r=0$) to the massive phase ($r > 0$) can be thought of as the spontaneous breaking of the $\mathcal{N}=1$ SUSY in 1+1-D \cite{kastor}. The associated  goldstino mode precisely corresponds to the free Majorana fermion formulation of the 1+1-D Ising critical point. This idea generalizes to the higher dimensional supersymmetric models considered here, including the 2+1-D boundaries of TSC and TI, where again the topological surface states are identified with the goldstino. This suggests an intriguing link between space-time SUSY and topological phases.

The transitions discussed in this paper have several unique, experimentally relevant signatures. As already discussed, at the critical point, supersymmetry  leads to precise relations between the scaling dimension of interacting fermion and boson modes, whose value, in some cases, can be determined exactly. As one moves away from the transition into the symmetry broken phase, the mass of the fermion, and the mass of the amplitude mode of the boson are equal to each other, again a consequence of supersymmetry. 

To conclude, we find emergent SUSY at transitions at the boundary of topological superconductors and insulators, using analytical field-theoretic arguments and numerical simulations. Potential routes to experimental realization were discussed. Finally, the ubiquitous appearance of SUSY at these surface topological transitions points to an intriguing connection between supersymmetry and topological phases, which remains to be explored.

{\it Acknowledgements:} We thank M. P. A. Fisher,  D. A. Huse, T. Mizushima, R. S. K. Mong and S. Trivedi for discussions. This work is supported by ARO MURI Grant W911-NF-12-0461 (AV), by the DOE Office of Basic Energy Sciences under grant DE-FG02-06ER46305 (DNS), and in part by the National Science Foundation under Grant No. NSF PHY11-25915 (TG).

\appendix

\section{Details of the lattice Model to realize D=1+1 SUSY critical point} \label{sec:1dmodeldetails}

We seek a lattice model in D =1+1 dimension, which yields the action $S^{1+1}$ in the main text at low energies. Although it is impossible to realize the edge of a D=2+1 topological superconductor in a purely D=1+1 system, we can construct an equivalent model where time reversal symmetry is traded for a different symmetry (here, a symmetry transformation that includes translation). As mentioned in the main text, the model is given by Eq.\ref{eq:lattice1d}:

\begin{eqnarray}
H &=&  -i\sum_j \left [ 1-g\mu^z_{j+\frac12}\right ]  \chi_j \chi_{j+1} + H_b \\ \nonumber
H_b &=& \sum_j   \left [{\mathcal J}\mu^z_{j-1/2}\mu^z_{j+1/2} - h  \mu^x_{j+1/2}\right ] 
%H_{\rm int} &=&  ig \sum_j \chi_j \mu^z_{j+\frac12}\chi_{j+1}\\ \nonumber
\end{eqnarray}

\noindent where $\chi_j$ is a single Majorana fermion at site $j$ while the Ising spins $\mu^z_{j+1/2}$ sit on bond centers. One may rewrite the Majorana fermions $\chi$ in terms of spin-1/2 spins which we denote by $\sigma$, via Jordan-Wigner transformation. Denoting $\mu_{j+1/2} = \mu_a$ for $j$ even and  $\mu_{j+1/2} = \mu_b$ for $j$ odd, the above model can therefore be rewritten entirely in terms of spins:

\be
H = H_1 + H_2 + H_3  \label{eq:H}
\ee

where

\bea
H_1 & = & -\sum_i \left(\sz_i \sz_{i+1} +  \sx_i\right) \label{H1spin} \nonumber \\
%& = & - i\sum_i \left (  \chi_i\overline{\chi}_{i}+\overline{\chi}_{i} \chi_{i+1}\right )\label{H1maj} \\
H_2 & = & \sum_i \left(\mathcal J \mz_{i,a} \mz_{i,b} + \mathcal J \mz_{i,b} \mz_{i+1,a} - h (\mx_{i,a} + \mx_{i,b}) \right)  \label{H2} \nonumber \\
%H_3 & = &  i g\sum_i \left(\chi_i \overline{\chi}_i \mu^z_{i,a} - \overline{\chi}_i \chi_{i+1} \mu^z_{i,b} \right) \\
H_3 & = &  g\sum_i \left(\sigma^x_i \mu^z_{i,a} - \sigma^z_i \sigma^z_{i+1} \mu^z_{i,b} \right) \nonumber \label{H3maj}
\eea

The above  models has an intricate symmetry, which we denote by $\mathcal{T'}$, that ensures that $\vec{\sigma}$ spins stay gapless throughout the whole phase when $\vec{\mu}$ spins are disordered. In particular, the
above model is invariant under:  $\sz_i \sz_{i+1} \xrightarrow{\mathcal{T'}}  \sx_{i+1}, \,
\sx_i \xrightarrow{\mathcal{T'}} \sz_i \sz_{i+1},\,
\mz_{i,a}  \xrightarrow{\mathcal{T'}}  -\mz_{i,b},\,
 \mz_{i,b}  \xrightarrow{\mathcal{T'}}  -\mz_{i+1,a} ,\,
\mx_{i,a}  \xrightarrow{\mathcal{T'}}  \mx_{i,b},\,
\mx_{i,b}  \xrightarrow{\mathcal{T'}}  \mx_{i+1,a} $. Physically, this symmetry exploits the fact that the transverse field Ising model is self-dual at the criticality. The self-duality interchanges the terms $-\sum_i \,\sz_i \sz_{i+1}$ and $- \sum_i \,\sx_i$ in $H_1$ and this is precisely the action of the symmetry $ \mathcal{T'}$ on the $\vec{\sigma}$ spins. Since the symmetry is respected by the full interacting Hamiltonian $H = H_1 + H_2 + H_3$, the spins $\vec{\sigma}$ remain gapless unless the $\mathcal{T'}$ symmetry is broken spontaneously. This happens when $\langle \mu_z \rangle \neq 0$, in the exact analogy with the action $S$ in Eqn.\ref{eq:lattice1d} where $\langle \phi \rangle$  provides mass gap to the helical Majorana modes. 

\section{Correlation functions and velocity anisotropy at the Tricritical Ising point} \label{sec:tricritcorr}

As discussed in the main text, SUSY imposes strong constraints on the correlation functions of various operators at the critical point.  To see how this arises, we first note that the action $S^{1+1}$ for the boundary of a 2D TSC is supersymmetric when $r = 0, v_\phi = 1$ and $u = \frac{g^2}{2}$ {\it(1,2)}. Indeed, if these constraints are met,  then the action is invariant under the following transformation {\it(2)}:

\bea
\delta \phi & = & \tilde{\epsilon} \chi  \label{fermirot1} \\
\delta \chi & = & ( \frac{-i\gamma_\mu  \partial_\mu \phi}{2} + i\frac{g}{4} \phi^2)\epsilon \label{fermirot2}
\eea

\begin{figure}
\begin{centering}
\vspace{-0.5in}
\hspace{-1in}
\includegraphics[scale=0.45]{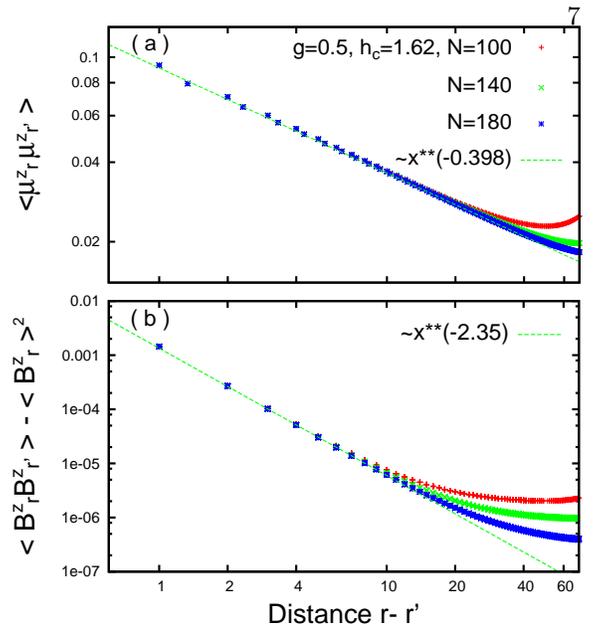}

\caption{The top figure shows the scaling of the correlation function $\langle \mu^z_r \mu^z_{r'} \rangle$ while the bottom one shows the bond-bond correlation function $\langle \mu^z_r \mu^z_{r+1} \mu^z_{r'} \mu^z_{r'+1} \rangle - \langle \mu^z_r \mu^z_{r+1}\rangle^2$ (we denote the bond operator $B^z_{r}=\mu^z_{r}\mu^z_{r+1}$ in the label of the figure) at the critical point between the two phases in Fig.2 in the main text. Emergent supersymmetry at the critical point implies that the difference in the power-law exponents for these two correlation functions is exactly two while the precise values of the exponents themselves are also found to be consistent with the tricritical Ising model (see the main text for details).} \label{fig5_super}
\par\end{centering}
\end{figure}

\noindent Such `fermionic rotational invariance' implies that the velocity as well as the anomalous dimension of the Majorana fermion $\chi$ equals that of the scalar $\phi$. Here  $\epsilon = [\epsilon_1\,\,\epsilon_2]^T$ is an arbitrary two-component Majorana variable,  $\tilde{\epsilon} = \epsilon\,i \sigma_y$, $\chi = [\chi_R \,\, \chi_L]$,  $\gamma_\tau = \sigma_y$,  $\gamma_x = \sigma_x$, and the Pauli matrices $\vec{\sigma}$ act on the spinor index of fermions $\chi$ and $\epsilon$. Of course, the numerical solution of our lattice model demonstrates that $\mathcal{N}=1$ supersymmetry emerges dynamically at the QPT by tuning only one parameter (in contrast to the  tuning of four parameters: $r, v_\phi, \frac{g^2}{u}$ and the Majorana fermion mass). In particular,

\bea
\langle \chi_L(r) \chi_L(0) \rangle & \sim & \langle \chi_R(r) \chi_R(0) \rangle \sim \frac{1}{r^{1+\eta_1}} \\
\langle \phi(r) \phi(0) \rangle & \sim & \frac{1}{r^{\eta_2}}
\eea

\noindent with $\eta_1 = \eta_2 = \eta$. It is known that $\eta = 0.4$ for the SUSY tricritical Ising point and we verified this using DMRG (see Fig.\ref{fig5_super}).  Supersymmetry also implies that the difference of the scaling dimension of the operators $\mu^z_i$ and $\mu^z_i \mu^z_{i+1}$ is precisely unity  {\it(1)}. Specifically {\it(1)}, $\langle \mu^z_r \mu^z_{r'} \rangle \sim \frac{1}{|r-r'|^{0.4}}$ while $\langle \mu^z_r \mu^z_{r+1} \mu^z_{r'} \mu^z_{r'+1}\rangle -\langle \mu^z_r\mu^z_{r+1}\rangle^2  \sim \frac{1}{|r-r'|^{2.4}}$. We verified this particular  prediction in our DMRG simulations and the results are shown in Fig.\ref{fig5_super}. Finally, as a further check of our numerical results, the scaling dimension of the operator $\sigma^z$ is close to the expected $3/80$ at the tricritical point.

Fig.\ref{sup1} shows the entanglement entropy at the critical point for the parameter $\mathcal{J} = 4$. We find that the critical point survives when for such a large anisotropy in the bare velocities of the boson and the fermion.

\section{Renormalization Group for Magnetic Instability of 3D TSC} \label{sec:3Drg}

The field theory for the transition is given by Eq.\ref{eq:fieldtheory} with $d=2$:

\begin{eqnarray} 
\nonumber
S^{2+1} &=& \int d\tau \,d^2x\, \,  [\underbrace{\frac12 \bar{\chi} \displaystyle{\not}\partial \chi  +\frac12 (\partial_\mu \phi)^2  +\frac r2 \phi^2}_{S_0} \\
&& + \underbrace{g\phi \bar{\chi}\chi  +u\phi^4}_{S_1}] 
\end{eqnarray}
where $\chi^T=(\chi_1,\,\chi_2)$ is a two component real fermion field with $\bar{\chi} = \chi^T\gamma^0 \,{\rm and}\, \displaystyle{\not}\partial =  \gamma^\mu\partial_\mu $ where $\gamma^0 = \sigma_y,\, \gamma^1=-\sigma_z,\,\gamma^2 = \sigma_x$ in terms of the Pauli matrices.

Above we have separated the quadratic part of the action ($=S_0$) from the interactions ($=S_1$). This theory has $\mathcal{N} = 1$ supersymmetry when $u = g^2/2$, and $r= 0$ {\it(2)}. We will demonstrate below that supersymmetry emerges dynamically in the above theory by tuning only the boson mass $\sqrt{r}$ to zero, even when the bare couplings $u$ and $g$ do not satisfy this relation. Note that here we have assumed that the bare boson and fermion velocities are equal. We will justify this assumption at the end by considering the RG flow of velocity difference between the boson and the fermion.

We now perform Wilson RG on the partition function

\be 
Z = \int D \phi D \chi e^{-S}
\ee

Following the standard procedure, we separate the fields into slow and fast components $\phi_{>},\phi_{<},\chi_{>},\chi_{<}$. The quadratic part $S_0$ decouples into fast and slow components without any mixing term, leading to:

\bea
Z & = & \int D\phi_< D\chi_{<} e^{-S^{<}_0} \int D\phi_>  D\chi_{>}   e^{-S^{>}_0} e^{-S_1} \nonumber \\
& = & Z^{>}_0 \int D\phi_<\,  D\chi_{<} e^{-S^{<}_0} \left<  e^{-S_1}\right>_>
\eea

where $Z^{>}_0 =  \int D\phi_> \, D\chi_{>} \, e^{-S^{>}_0}$ and

\be
\left<  e^{-S_1}\right>_> = \frac{1}{Z^{>}_0}\int D\phi_> \, D\chi_{>} \, e^{-S^{>}_0} e^{-S_1}
\ee

The integration over the fast modes $\phi_>, \chi_{>}$ would modify the action for the slow modes  $\phi_<, \chi_{<}$, thus yielding the RG equations so desired. To control the RG flow, we perform an $\epsilon$-expansion where $\epsilon = 4-D$. Symmetry analysis suggests that if a stable fixed point exists, then $u^{*} = O(\epsilon)$ and $g^{*} = O(\sqrt{\epsilon})$. Therefore, a cumulant expansion in powers of $u, g$ is a legitimate option. We will need an expansion upto $O(g^4, u^2)$:

\bea
\left<  e^{-S_1}\right>_> & = &e^{-\left[ \left< S_1\right>_> - \frac{1}{2} \left< S^2_1\right>_>  + \frac{1}{6} \left< S^3_1\right>_>  - \frac{1}{24} \left< S^4_1\right>_> \right]}
\eea

\noindent where we have dropped the terms such as $\langle S_1\rangle_{>}^2$ that do not contribute.
We next write down the contribution from each of
 the four terms in the cumulant expression to the renormalization of the total action $S$.

\vspace{0.5cm}

$\boxed{O(g,u):\,\left<S_1\right>_>}$

\bea
\left<S_1\right>_> = g \, \int d^Dr\,\, \chi_{<}^T (r) \sigma_y \chi_{<}(r) \, \phi_<(r) + u \int d^Dx \,\, \phi^4_{<}(r)
\eea
where $D = 2+1 = 3$ is the space-time dimension and $r = \left(\tau, x,y \right)$. Consider the elementary coarse-graining transformation $r' = r/s$. Under this transformation, $\chi_{<}(r) = \xi_\chi \chi(r')$ and
$\phi_<(r) = \xi_\phi \phi(r')$. $\xi_\chi$ and $\xi_\phi$ are related to the scaling dimensions $\Delta_\chi, \Delta_\phi$ of the fields $\xi$ and $\chi$
for $D = 4-\epsilon$ via $\xi_{\chi} = s^{-\Delta_\chi} = s^{-\eta_\chi/2}$ and $\xi_{\phi} = s^{-\Delta_\phi} = s^{-(d-1+\eta_\phi)/2}$ where $\eta_\chi$ and $\eta_\phi$ are the anomalous dimension of the fermion and boson modes respectively. Re-expressing $\left<S_1\right>_>$ in terms of rescaled variables, one finds $g(s) = g s^{\epsilon/2-\eta_\chi-\eta_\phi/2}$ and $u(s) = u s^{\epsilon - 2 \eta_\phi} $ at this order.

$\boxed{O(g^2,u^2):\,- \frac{1}{2} \left< S^2_1\right>_> }$

These terms lead to an  anomalous dimension  for the Majorana fermions $\chi$ and the boson $\phi$, and also renormalizes the interaction strength $u$ between bosons.

\textbf{Anomalous dimension $\eta_\chi$ of the Majorana fermion:}

The relevant term is 

\be 
\Delta S = \frac{g^2}{2} \int d^D r d^D r' \langle \chi^T(r) \sigma_y \chi(r) \,\phi(r) \chi^T(r') \sigma_y \chi(r') \,\phi(r') \rangle_{>} 
\ee 
which can be shown to contribute 

\bea
\Delta S & = & \int d^D p\,\,\, 4 \times  \frac{g^2}{2} \times \left( \chi_{<}^T i\sigma_y p_\mu \gamma_\mu \chi_{<} \right)\times \frac{N_d}{2} \log(\Lambda/p) \nonumber
\eea
where $N_D = \frac{A(S_{D-1})}{(2\pi)^D}$ with $A(S_{D-1})$ as the area of the $D-1$ sphere; $\gamma_\tau = \sigma_y, \gamma_x = \sigma_z, \gamma_y = -\sigma_x$. Therefore the anomalous dimension of Majorana $\eta_\chi = 2 g^2 N_D$.

\textbf{Anomalous dimension $\eta_\phi$ of the scalar:}

Similarly, one finds a contribution 

\bea
\Delta S & = &  \int d^D p \,\,\frac{g^2   N_D  |\phi_{<}(p)|^2 p^2 \log(p)}{2} \times  \textrm{Trace}(\mathbb{I})  \nonumber
\eea
to the anomalous dimension $\eta_\phi$ of the bosons $\phi$. We set $ \textrm{Trace}(\mathbb{I}) = 2$ because under our scheme the spinors live in three space-time dimensions.  This implies that $\eta_\chi = 2g^{*2} N_D = \eta_\phi$ where $g^{*}$ is the fixed point value of the coupling $g$ that we determine below.

\textbf{Renormalization of $u$ at $O(u^2)$:}

One finds a contribution

\bea
\Delta S & = &  -36 u^2 N_D \log(s) \int d^D r {\phi^4}_<(r) 
\eea

$\boxed{O(g^3):\,\frac{1}{6}
\left< S^3_1\right>_>  }$

\textbf{Renormalization of $g$ at $O(g^3)$:}

Only the term $\left<S^3_1\right>_>$ contributes to the renormalization of the interaction $g$ at this order. One finds

\bea
\Delta S & = &  -4 g^3 N_D \log(s) \int d^D r \,\,\chi^T_{<}(r) \sigma_y \chi(r) \, \phi_<(r) 
\eea

$\boxed{O(g^4):\,\frac{-1}{24} \left< S^4_1\right> }$ 

\textbf{Renormalization of $u$ at $O(g^4)$:}

\bea
\Delta S & = &  4 g^4 N_D \log(s) \int d^D r {\phi^4}_<(r) 
\eea

\subsection{Renormalization Group Equations and Fixed Point}

Putting everything together and setting $s = e^{dl} \approx 1 + dl$, the renormalization of coupling $g$ is given by,

\bea
g(1+dl) = \left(1 + dl\left(\epsilon/2 - 3g^2 N_D \right)\right)\left( g - 4 g^3 N_D dl\right) \nonumber
\eea

Or,

\be \boxed{
\frac{dg^2}{dl} = \epsilon g^2 - 14 g^4 N_D }
\ee

Similarly,

\be 
u(1+ dl) = \left( 1 + (\epsilon - 4 g^2 N_D) dl\right)\left( u - 36 u^2 N_D dl + 4 g^4 N_D dl \right) \nonumber
\ee

Or, 

\be
\boxed{
\frac{du}{dl} = \epsilon u - 4 g^2 u N_D - 36 u^2 N_D + 4 g^4 N_D }
\ee

Solving the above equations, one finds

\be
(g^2)^{*} = \frac{\epsilon}{14N_D}, \,\,\, u^{*} =  \frac{\epsilon}{28N_D}
\ee
which precisely satisfies the condition $u^{*} = \frac{{g^2}^{*}}{2}$ required for supersymmetry. Substituting these fixed-point values into the expression for anomalous dimensions, one finds

\be
\boxed{
\eta_\chi  =  \eta_\phi = \frac{\epsilon}{7}}
\ee

which means that $\eta_\chi  =  \eta_\phi = \frac{1}{7}$ at $O(\epsilon)$.

\subsection{Emergence of Lorentz Invariance in the Non-Relativistic System}
\label{Sec:SUSYvelocity}
Let us consider the effect of different bare velocities for the fermion and the boson. That is, we consider the action

\bea
S & = & \int d^Dr \,\,\frac{1}{2} \chi^T(\partial_\tau - i v_\chi \sigma_x \partial_x - i v_\chi \sigma_z \partial_y)\chi + \nonumber \\ & &  \frac{(\partial_\tau \phi)^2}{2} + v^2_\phi \frac{(\partial_x \phi)^2 + (\partial_y \phi)^2}{2}+ \frac{r \phi^2}{2}  + \nonumber \\ & &  u \phi^4 + g \chi^T \sigma_y \chi \,\phi
\eea

The velocities will receive renormalization from the second order cumulant.

\textbf{Boson Velocity Renormalization:}

The relevant term in the renormalized action is

\bea 
\Delta S & = & \frac{1}{v^3_\chi}\int d^D p \,\,2g^2   N_D  |\phi_{<}(p)|^2 \frac{(p^2_0 + v^2_\chi \vec{p}^2) \log(\Lambda/p)}{2} \nonumber
\eea

This implies that the boson velocity renormalizes as
\be 
v'^2_\phi = v^2_\phi \left( 1 + \frac{2g^2N_D}{v^3_\chi}\left( \left(\frac{v_\chi}{v_\phi}\right)^2-1\right) \log(\Lambda) \right)
\ee

Or,

\be 
\boxed{
\frac{dv_\phi}{dl} = \frac{g^2N_D(v^2_\chi-v^2_\phi)}{v_\phi v^3_\chi}}  \label{eq:cb_rg}
\ee

\textbf{Fermion Velocity Renormalization:}

In this case, the relevant term is 

\bea 
\Delta S & = & 2g^2 \int d^Dk \,\,\,\chi^T (k) i \sigma_y \times \nonumber \\ & & \int \frac{d^Dp \,\,\,(p_0+k_0)\gamma_0 + v_\chi (\vec{p} + \vec{k}).\vec{\gamma}}{(2\pi)^D((p_0+k_0)^2 + v^2_\chi(\vec{p} + \vec{k})^2)(p^2_0 + v^2_\phi \vec{p}^2)} \chi(-k) \nonumber \\
& = & I_1 + I_2
\eea

where 
\bea 
I_1 & = & 2g^2  \int d^Dk \,\,\, \chi^T  i \sigma_y (k_0 \gamma_0 + v_\chi \vec{k}.\vec{\gamma}) \chi \times \nonumber \\ & &  \int \frac{d^Dp}{(2\pi)^D(p^2_0 + v^2_\phi \vec{p}^2)(p^2_0 + v^2_\chi \vec{p}^2)} \nonumber
\eea 

and 

\bea 
I_2 & = & 2g^2 \int d^Dk \,\,\, \chi^T  i \sigma_y \times  \nonumber \\ & &  \int \frac{d^Dp \,\,\, p_0\gamma_0 + v_\chi \vec{p}.\vec{\gamma}}{(2\pi)^D((p_0+k_0)^2 + v^2_\chi(\vec{p} + \vec{k})^2)(p^2_0 + v^2_\phi \vec{p}^2)}  \chi \nonumber
\eea

Putting everything together, one finds

\be
\Delta S = \frac{g^2 N_d}{v^3_\chi} \int d^Dk \,\,\, \log(\Lambda/k)  \chi^T  i \sigma_y  \left(f_0 k_0 \gamma_0 + f_1 \vec{k}.\vec{\gamma}  \right) \chi 
\ee
with $f_0 = \frac{4}{\alpha(1+\alpha)^2}$ and $f_1 = \frac{4(2\alpha+1)}{3\alpha(1+\alpha)^2}$ with $\alpha = v_\phi/v_\chi$. This leads to the following RG flow for $v_\chi$:

\be 
\boxed{
\frac{dv_\chi}{dl} = \frac{16g^2N_D(v_\phi-v_\chi)}{3v_\phi (v_\phi+v_\chi)^2}}  \label{eq:cf_rg}
\ee

Denoting the velocity difference as $\Delta v = v_\phi - v_\chi$, and setting $v_\phi \approx v_\chi = v$, one may combine the above equation with the one for the flow of $\Delta v$. One finds,

\bea
\frac{d\,\Delta v}{dl} & = & -\frac{10g^2 N_D}{3 v^3} \Delta v \nonumber \\
& \approx &  -\frac{5 \epsilon}{21 v^3} \Delta v
\eea

\noindent near the transition. This imply that $v_\phi = v_\chi$ at the fixed point, and therefore our starting point in the previous section, where we assumed Lorentz invariance, is justified.
\end{document}